\documentstyle[epsfig,prl,floats,aps]{revtex}
\begin{document}
\input psfig.sty
\twocolumn
\title{Dynamical Response of a Bose-Einstein Condensate to a
  Discontinuous Change in Internal State}  
\author{M.R. Matthews, D.S.
  Hall, D.S. Jin,\cite{eric} J.R.  Ensher, C.E. Wieman, and E.A.~Cornell\cite{eric}}
\address{JILA and Department of Physics,
  University of Colorado and National Institute of Standards and Technology, Boulder, Colorado 80309-0440} 
\author{F. Dalfovo, C. Minniti, and S. Stringari}
\address{ Dipartimento di Fisica, Universit\`{a} di Trento, and Istituto
  Nazionale Fisica della Materia, I-38050 Povo, Italy}
\date{March 2, 1998}
\maketitle
\begin{abstract}
  A two-photon transition is used to convert an arbitrary fraction of
  the ${}^{87}\mathrm{Rb}$ atoms in a $\left |F=1,m_f=-1 \right >$
  condensate to the $\left |F=2,m_f=1 \right >$ state.  Transferring
  the entire population imposes a discontinuous change on the
  condensate's mean-field repulsion, which leaves a residual ringing
  in the condensate width.  A calculation based on Gross-Pitaevskii
  theory agrees well with the observed behavior, and from the
  comparison we obtain the ratio of the intraspecies scattering
  lengths for the two states, $a_{\left | 1, -1 \right >}/ a_{\left
      |2,1 \right >}$=1.062(12).
\end{abstract}
\par 
The effects of interactions on dilute atomic Bose-Einstein condensates
(BEC)\cite{1stbec} have been studied in several
contexts, including density and momentum distributions
\cite{exp}, collective excitations
\cite{excite}, specific heat \cite{jilaheat}, speed of
sound \cite{mitsound}, and limited condensate number\cite{ricenum}.
There has been excellent quantitative agreement between these
experiments and theory, made possible by both experimental advances
and the fact that the interactions can be modeled relatively simply.
Despite complicated interatomic potentials, the mean-field interaction
in condensates is well-characterized by a single parameter, the
$s$-wave scattering length $a$.  Previous quantitative experiments on
interactions have all been done using single-component condensates
with a constant value of $a$, but several authors have proposed using
external optical or magnetic fields to shift the mean-field
interaction by perturbing the interatomic potential
\cite{da1,da2,da3}.
\par
We present a method for the creation of condensate mixtures using
radiofrequency (rf) and microwave fields.  We are able to transfer
abruptly a trapped condensate of one hyperfine state into a coherent
superposition of two trapped hyperfine states, and then watch the
subsequent dynamical behavior.  This approach makes possible a variety
of two-species BEC studies.  In this paper, we examine quantitatively
one special case, in which all condensate atoms are converted from one
state to another.  Since these two states have slightly different
values of $a$, the sudden change in self-interaction gives rise to
oscillatory spatial behavior of the condensate wave function
\cite{shlyap}.  The scattering length ratio can be extracted from a
model using analytical equations of motion
for the condensate widths \cite{shlyap,castin,dalf} based on Gross-Pitaevskii
theory.
\par
The first demonstration of a binary mixture of condensates by Myatt
\emph{et al}.\cite{1stdouble} produced overlapping condensates of the
$5S_{1/2}\left |F=1,m_f=-1 \right >$ and $\left |F=2,m_f=2 \right >$
states of ${}^{87}\mathrm{Rb}$.  The ratio of the magnetic moments of
these states is 1:2, so the condensates experience different
potentials in a magnetic trap and are displaced unequally from the
trap center by gravity.  Due to an accidental coincidence between the
singlet and triplet scattering lengths of ${}^{87}\mathrm{Rb}$,
collisional loss is reduced and any mixture of spin states will be
relatively long-lived \cite{1stdouble,life1,life2}.  Here, we use
mixtures of $\left | 1, -1 \right >$ and $\left |2,1 \right >$ states,
which possess several advantages.  First, these two states have
essentially identical magnetic moments, and hence feel identical
confining potentials.  Second, one can conveniently and quickly change
atoms from the $\left | 1, -1 \right >$ state to the $\left |2,1
\right >$ state by a two-photon transition (microwave plus rf).
Finally, we can selectively image the different components using
appropriately tuned lasers.
\par
The apparatus uses a multiply loaded, double magneto-optical trap
(MOT) scheme \cite{2mots}, consisting of two MOT regions connected by
a 60 cm long, 1 cm diameter transfer tube. In the first MOT, ${}^{87}\mathrm{Rb}$
atoms are collected from a background Rb vapor, produced by heating a
Rb getter source \cite{getter}.  The trapping beams are then turned
off and a near-resonant laser beam is pulsed on to push the atoms
through the transfer tube into the second MOT.  To prevent collisions
with the walls, the outside of the transfer tube is lined with strips
of permanent magnetic material which form a hexapole guiding field.
The atoms exit the tube and are captured by the second MOT, which has
a 70 second lifetime due to differential pumping.  This process is
repeated twice a second for $\sim$10~s until approximately $10^9$
atoms have accumulated in the second MOT.
\par 
The atoms are further cooled in an optical molasses and then optically
pumped into the $\left | 1, -1 \right >$ state.  These atoms are then
transferred into a time-averaged, orbiting potential (TOP) magnetic
trap \cite{top}.  The TOP trap consists of a quadrupole magnetic field
with an axial gradient of $\sim$250~G/cm and a uniform magnetic field
which rotates at 1.8~kHz.  The fraction of atoms captured from
molasses in the TOP trap depends approximately linearly on the
magnitude of the rotating field.  We use the maximum rotating field
achievable in our apparatus (49~G) to capture up to 50\% of the atoms
from the MOT, substantially enhancing the initial number as compared
to our previous setup \cite{1stbec}.  As in \cite{1stbec}, the
confined cloud is then magnetically compressed and evaporatively
cooled to a temperature where there is no noticeable non-condensed
fraction remaining, corresponding to a temperature less than 0.4 of
the BEC transition temperature.  Evaporative cooling requires about 30
seconds, and typically produces a condensate of $5\times 10^5$ atoms.
\par 
The magnetic moments of the $\left | 1, -1
  \right >$ and $\left |2,1 \right >$ states are nearly the same
to first order in magnetic field.  Second-order dependence, the
nuclear magnetic moment, and small effects due to the time varying
nature of the TOP trap must be taken into account when calculating the
exact trap potential \cite{timetop,bohn}. The quadrupole and rotating
magnetic fields are adiabatically changed after evaporation to make
the trap potentials the same (to within 0.3\%) for the two states.
Evaporation takes place in a trap with radial frequency $\nu_r=35$~Hz;
subsequent work occurs at $\nu_r=17$~Hz with a rotating field
magnitude of 5.7~G ($\nu_r=\nu_z/\sqrt{8}$
for the TOP trap).
\par
\begin{figure}[t]
\begin{center}
\psfig{figure=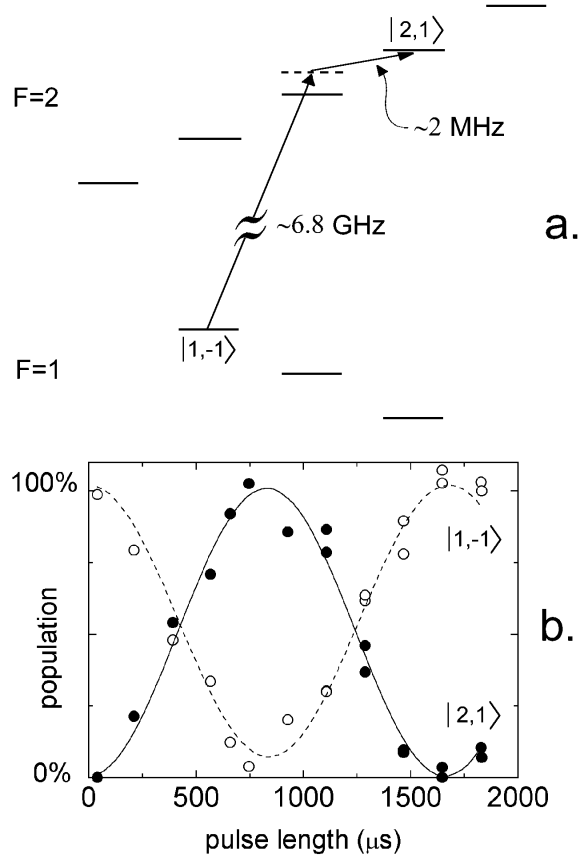,width=0.95\linewidth,clip=}
\end{center}
\caption{(a). A diagram of the ground state hyperfine
  levels ($F=1,2$) of ${}^{87}\mathrm{Rb}$ shown with Zeeman splitting
  due to the
  presence of a magnetic field.  The two-photon transition is driven
  between the $\left | 1, -1
  \right >$ and $\left |2,1 \right >$ states.  (b) The Rabi oscillation
  of population between the $\left | 1, -1
  \right >$ (open circles) and $\left |2,1 \right >$ 
  states (solid circles) as a function of the two-photon drive
  duration.  The lines are fit to the data and show the expected
  sinusoidal oscillation.}
\label{fig1}
\end{figure}

The two-photon transition used to change the hyperfine state is shown
schematically in Fig.\ \ref{fig1}a.  We apply a pulse of microwave
radiation at a frequency slightly less than the ground state hyperfine
splitting of ${}^{87}\mathrm{Rb}$ ($\sim$6.8~GHz) along with a $\sim$2~MHz
rf magnetic field.  This connects the $\left | 1, -1
  \right >$ state to the $\left |2,1 \right >$
state via an intermediate virtual state with a detuning of
2.2~MHz from the $\left | 2, 0 \right >$ state .  The two-photon
Rabi frequency is 600~Hz, which is much faster than the characteristic
frequency $2\nu_z=96$~Hz for the condensate to change shape.

\par
We are able to view either component of a mixture of $\left | 1, -1
\right >$ and $\left |2,1 \right >$ condensates or the combined
density distribution.  Observation of the atoms is a destructive
measurement in which the TOP trap is suddenly turned off and the
condensate is left to expand ballistically for 22~ms.  To image the
combined density distribution of both states, a short pulse of
``repump'' light pumps the atoms from the $\left | 1, -1 \right >$
state into the $F=2$ manifold.  About $100\mu s$ later, a $\sigma^+$
circularly polarized probe beam 17~MHz detuned from the $5S_{1/2}$
$F=2$ to $5P_{3/2}$ $F'=3$ cycling transition is scattered by the
atoms and the shadow is imaged onto a charge-coupled device camera.
Imaging only the $\left |2,1 \right >$ condensate uses the same
procedure, except that the pulse of repump light is omitted.  Atoms in
the $\left | 1, -1 \right >$ state are far (6.8 GHz) from resonance
and invisible to the probe beam.  Imaging only the $\left | 1, -1
\right >$ atoms is similar to viewing both species simultaneously, but
the $\left |2,1 \right >$ atoms are first ``blown away'' from the
imaging region by a 2~ms, 60 $\mu$W/cm$^2$ pulse of $F=2$ to $F'=3$
light applied near the end of the ballistic expansion.  This light has
no effect on the $\left | 1, -1 \right >$ atoms, to which a subsequent
pulse of repump light is applied, pumping them into the $F=2$
manifold for probing.  In Fig.\ \ref{fig1}b we show the population of
each state as a function of the two-photon pulse length.  The Rabi
oscillations have nearly 100\% contrast, indicating that we are able
to put any desired fraction into the $\left |2,1 \right >$ state.

\par
\begin{figure}
\begin{center}
\psfig{figure=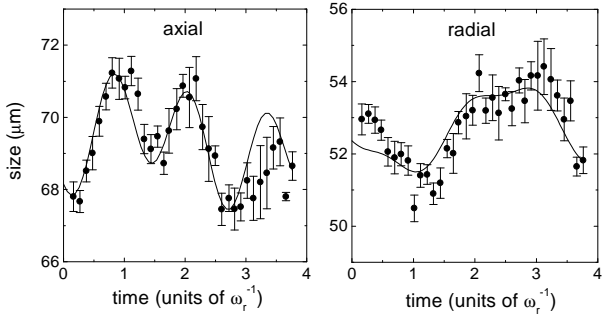,width=\linewidth,clip=}
\end{center}
\caption{Oscillation in the width of the cloud in both the axial and
  radial direction due to the instantaneous change in scattering
  length.  The widths are for condensates as a function of free evolution time in units of
  the radial trap period ($\omega_r^{-1}=9.4$ ms), followed by 22~ms
  of ballistic expansion.  Each point is the average of approximately
  10 measurements.  Note that the fractional change in width is quite small.
  The solid line is a fit of the model to the data, with only the
  amplitude of the oscillation and the initial size as free parameters.}
\label{fig2}
\end{figure}

We have investigated the dynamical behavior of the condensate after
more than $99.5$\% of the atoms are transferred from the $\left | 1,
  -1 \right >$ state to the $\left |2,1 \right >$ state.  After the
transfer, the condensate is allowed to evolve in the trap for a time
interval $T$ before the cloud is released from the trap and probed.
The experiment is repeated for different values of $T$, and the axial
and radial widths of the condensate are measured using a fit to the
two-dimensional condensate image \cite{TFfit}.  Fig.\ \ref{fig2}
illustrates the time dependence of both the axial and radial size of
the atom cloud.  Qualitatively, the data for both dimensions are
consistent with a ``compression'' oscillation.  The condensate shrinks
for very early times, indicating a weaker mean-field repulsion in the
$\left |2,1 \right >$ state.

\par
In order to describe the response to the change in scattering 
length, we use the Gross-Pitaevskii (GP) equation for the 
condensate wave function, with a time-dependent interaction term:
\begin{equation}
i \hbar { \partial \Phi ({\bf r},t) \over \partial t} = 
\left[ - {\hbar^2 \nabla^2 \over 2m} + V({\bf r}) +
U(t) | \Phi ({\bf r},t)|^2 \right] \Phi ({\bf r},t) \; ,
\end{equation}
where $V({\bf r})$ is the confining potential, $m$ is the 
$^{87}$Rb mass, and $U(t)= 4\pi \hbar^2 a(t)/m$.
This has the form of a nonlinear Schr\"odinger equation, with 
a mean-field characterized by the $s$-wave scattering 
length $a(t)$.  For $t<0$, $a(t)=a_1$, the scattering length for 
$|1,-1\rangle$ atoms on $|1,-1\rangle$ atoms. For 
$t>0$, $a(t)=a_2$, the $|2,1\rangle$ on $|2,1\rangle$ 
scattering length. The condensate is formed in its ground-state 
in which the density profile  $n({\bf r})=|\Phi({\bf r})|^2$ 
is constant in time, and with the axial and radial widths $(w_z,w_r)$ 
determined by $V({\bf r})$, $a_1$, and the number of atoms $N$. 
After a sudden change in $a(t)$, $\Phi({\bf r},t)$ is projected 
onto a coherent superposition of its new ground-state and collective
vibrational modes. Instead of solving the complete GP equation,
one can use the Thomas-Fermi (TF) approximation, which corresponds
to neglecting the quantum pressure term in the kinetic
energy of the condensate. In this case, one can replace the GP 
equation  with a pair of scalar equations of motion for the 
condensate widths \cite{shlyap,castin,dalf}:
\begin{equation}
\ddot{w}_{i} + (2\pi \nu_{i})^2 w_{i} - 
\left( {15 \hbar^2 N a(t) \over 7^{5/2} m^2 } \right)
{ 1 \over w_r^2 w_z w_{i} } = 0 \; .
\label{eq2} 
\end{equation}
Here the widths $w_{r}=\sqrt{\langle r^2 \rangle }$ and $w_{z}
=\sqrt{\langle z^2 \rangle }$ are the radial and axial rms radii,
and the subscript $i$ is to be replaced with either $r$ or $z$ for the 
respective widths.  The TF approximation holds when $N$ is large \cite{pg}
and, in this limit, the GP theory coincides with the hydrodynamic theory
of superfluids \cite{strin}.  
\par
Time-dependent behavior is predicted by numerically integrating
Eqs.~(2). The initial conditions ($t<0$) are that the widths are at
the values determined by the ground-state solution,
$\dot{w}_r=\dot{w}_z=0$, and $a(t)=a_1$. After $t=0$, $a(t)=a_2$ and
the integration proceeds for time $T$. We model the removal of the
trap by setting $\nu_{r}=\nu_{z} =0$ and continue the numerical
integration for the $22$ ms free expansion. The axial and radial
widths subsequently expand with different speeds. In the TF
approximation, the ratio between the two expanded widths, averaged
over $T$, can be shown to depend only on $\nu_{r},\nu_{z}$ and the
expansion time \cite{dalf}.  In our case, the theory predicts
$w_z/w_r=1.29$, in good agreement with the measured value
$w_z/w_r=1.31$. The oscillations of the widths correspond to a
superposition of two $m=0$ modes. Since no angular momentum is
imparted to the condensate by the change of scattering length, these
are the only modes excited.  The calculated mode frequencies are
$\nu/\nu_r=1.80$ and $\nu/\nu_r=4.99$, and turn out to be independent
of the amplitude in the range considered here. The two modes
contribute to the oscillations in $z$ and $r$ with a different phase and
amplitude, the axial motion behaving mainly as the fast mode and the
radial motion as the slow one.
\par
In each direction, the ratio of the amplitudes of the two modes is
predicted to be constant over the range of possible scattering lengths
relevant to the experiment. Thus, only the initial size and an overall
amplitude are used as adjustable parameters in comparison of theory
with experiment. The solid lines in Fig.\ \ref{fig2} show the
theoretical prediction using the best fit value of the oscillation
amplitude, which is related to the scattering length ratio $a_1/a_2$.
The predicted oscillations agree remarkably well with the shape of the
data, for both the frequency and the phase of the two modes. From the
fit amplitude we obtain the ratio of the scattering lengths
$a_1/a_2=1.062\pm 12$, which is consistent with the ratio of
$1.059_{-6}^{+4}$ obtained in a theoretical calculation of
binary collision parameters~\cite{burke}.

\par
\begin{figure}
\begin{center}
\psfig{figure=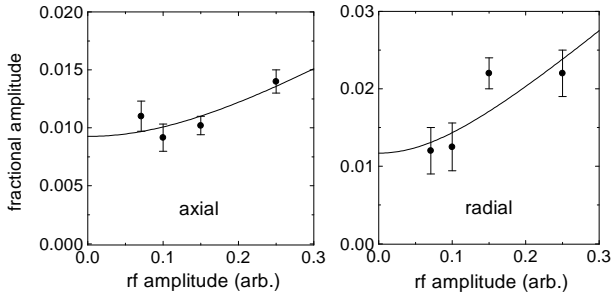,width=\linewidth,clip=}
\end{center}
\caption{Dependence of the oscillation amplitude on the
  rf strength in the two-photon transition.  The vertical axis
  represents the fitted value of the fractional excitation amplitude,
  with error bars from uncertainty in the fit.
  The solid line shows extrapolation to zero rf amplitude.}
\label{fig3}
\end{figure}

Analysis of the data is complicated slightly by an observed systematic
dependence of the oscillation amplitude on the magnitude of the rf
drive during the two-photon pulse.  We believe that this effect is
partly due to a small dressing of the atoms to the $\left |1,0\right
>$ or $\left |2,0\right > $ state by the rf, and partly due to a coupling
of the rf onto electronics controlling the trap potential.  Both
result in a change in strength of the confining potential during the
two-photon pulse which returns to normal when the pulse is complete.
After this impulsive perturbation, the BEC can be thought of as freely
evolving with an initial ``velocity'' in the width
($\dot{w_r},\dot{w_z} \neq 0$).  This is in contrast to the discrete
change in scattering length, in which the BEC width has an initial
offset from the equilibrium width, but no initial velocity.  The
systematic is manifested as an initial offset in the phase of the
resulting oscillation, which is indeed observed for large rf
amplitudes.  Our result for the ratio was obtained by extrapolating to
zero rf amplitude with a quadratic fit as shown in Fig.\ \ref{fig3}.

\par
We have explored the dynamical response of a Bose-Einstein condensate
to a sudden change in the interaction strength.  The agreement with
the model is excellent, and presents a new and precise way of
measuring a ratio of scattering lengths.  This also allows a test of
the theory for pressure shifts in atomic clocks \cite{clock}.  Our method of creating two fully
interpenetrating condensates will allow coherent control of the relative
population and detailed examination of spatial and phase dynamics in
the future.
\par
We acknowledge useful conversations with the other members of the JILA
BEC collaboration, and in particular with Chris Greene.  This work is
supported by the ONR, NSF, and NIST.


\begin{references}
\bibitem{TFfit} The expanded condensates are fit
  using a 2-dimensional projection of the Thomas-Fermi form for the
  density distribution in a harmonic trap.  The fit also provides us with
  a measure of the number of atoms: we have measured a lifetime of 300
  ms for the $\left |2,1 \right >$ state.  This gives an upper limit on the
  $\left |2,1 \right >$ inelastic loss rate of $3\times 10^{-13}$cm$^3$/s.  Number,
  and hence width is a decaying
  function of time, so we apply a small correction to the data
  (Fig.~\protect\ref{fig2}) based on this.
\end{references}
\end{document}